\documentclass[aps,pre,reprint,floatfix,amsmath,amssymb]{revtex4-1}

\usepackage{epsfig}
\usepackage{graphicx}
\usepackage{dcolumn}
\usepackage{bm}
\usepackage[colorlinks=true,dvipdfm]{hyperref}
\usepackage{amssymb}

\begin{document}
\title{Universal scaling in first-order phase transitions mixed with nucleation and growth}

\author{Fan Zhong}
\affiliation{State Key Laboratory of Optoelectronic Materials and
Technologies and School of Physics, Sun Yat-sen
University, Guangzhou 510275, People's Republic of China}

\date{\today}

\begin{abstract}
Matter exhibits phases and their transitions. These transitions are classified as first-order phase transitions (FOPTs) and continuous ones. While the latter has a well-established theory of the renormalization group, the former is only qualitatively accounted for by classical theories of nucleation, since their predictions often disagree with experiments by orders of magnitude. A theory to integrate FOPTs into the framework of the renormalization-group theory has been proposed but seems to contradict with extant wisdom and lacks numerical evidence. Here we show that universal hysteresis scaling as predicted by the renormalization-group theory emerges unambiguously when the theory is combined intimately with the theory of nucleation and growth in the FOPTs of the paradigmatic two-dimensional Ising model driven by a linearly varying externally applied field below its critical point. This not only provides a new method to rectify the nucleation theories, but also unifies the theories for both classes of transitions and FOPTs can be studied using universality and scaling similar to their continuous counterpart.
\end{abstract}

\maketitle

\section{\label{intro}Introduction}
Matter as a many-body system exists in various phases or their coexistence and its diversity comes from phase changes. It thus exhibits just phases and their transitions. These transitions are classified as first-order phase transitions (FOPTs) and continuous ones~\cite{Fisher67}, the latter including second and higher orders. Whereas the phases can be studied by a well-established framework and the continuous phase transitions have a well-established theory of the renormalization group (RG) that has predicted precise results in good agreement with experiments~\cite{Barmatz}, the FOPTs gain a different status in statistical physics.

FOPTs proceed through either nucleation and growth or spinodal decomposition~\cite{Gunton83,Bray,Binder2}. Although classical theories of nucleation~\cite{Becker,Becker1,Becker2,Zeldovich,books,books1,books2,Oxtoby92,Oxtoby921,Oxtoby922,Oxtoby923} and growth~\cite{Avrami,Avrami1,Avrami2} correctly account for the qualitative features of a transition, an agreement in the nucleation rate of even several orders of magnitude between theoretical predictions and experimental and numerical results is regarded as a feat~\cite{Oxtoby92,Oxtoby921,Oxtoby922,Oxtoby923,Filion,Filion1,Filion2}. A lot of improvements have thus been proposed and tested in the two-dimensional (2D) Ising model whose exact solution is available. One theory of nucleation, called FT hereafter, considers field theoretic corrections to the classical theory~\cite{Langer67,Gunther,Prestipino}. Upon being combined with Avrami's growth law~\cite{Avrami,Avrami1,Avrami2}, it was shown to accurately produce the results of hysteresis loop areas obtained from Monte Carlo simulations at a temperature $T$ below the critical temperature $T_c$ both in a single droplet (SD) and a multidroplet (MD) regimes---differentiated by whether only one or many droplet nucleate~\cite{Rikvold94}---even in the case of a sinusoidally varying applied external field $H$ with an adjustable parameter~\cite{Sides98,Sides99,Ramos,Zhongc}. Another theory, referred to as BD below, adds appropriate corrections to the droplet free energy of Becker and D\"{o}ring's nucleation theory~\cite{Becker2}. It was found to accurately predict nucleation rates for the 2D Ising model without adjustable parameters~\cite{Shneidman99,Ryu,Ryu1}.

However, it is well-known that classical nucleation theories are not applicable in spinodal decompositions in which the critical droplet for nucleation is of the size of the lattice constant and thus no nucleation is needed~\cite{Gunton83}. In contrary to the mean-field case, for systems with short-range interactions, although sharply defined spinodals that divide nucleation and growth from spinodal decomposition do not exist~\cite{Gunton83,Bray,Binder2}, one can nevertheless assume existence of fluctuation shifted underlying spinodals called ``instability'' points. Expanding around them below $T_c$ of a usual $\phi^4$ theory for critical phenomena then results in a $\phi^3$ theory for the FOPT due to the lack of the up--down symmetry in the expansion~\cite{Zhongl05,zhong16}. An RG theory for the FOPT can then be set up in parallel to the RG theory for the critical phenomena, giving rise to universality and dynamic scaling characterized by analogous ``instability'' exponents. The primary qualitative difference is that the nontrivial fixed points of such a theory are imaginary in values and are thus usually considered to be unphysical, though the instability exponents are real. Yet, counter-intuitively, imaginariness is physical in order for the $\phi^{3}$ theory to be mathematically convergent, since the system becomes unstable at the instability points upon renormalization and analytical continuation is necessary~\cite{Zhonge12}. Moreover, the degrees of freedom that need finite free energy costs for nucleation are coarse-grained away with the costs, indicating irrelevancy of nucleation to the scaling~\cite{Zhonge12}. Although no clear evidence of an overall power-law relationship was found for the magnetic hysteresis in a sinusoidally oscillating field in two dimensions~\cite{Thomas,Sides98,Sides99} in contrast to previous work~\cite{Rao,Char,Liang16}, recently, a dynamic scaling near a temperature other than the equilibrium transition point was again found numerically for the cooling FOPTs in the 2D Potts model with properly logarithmic corrections~\cite{Pelissetto16}. However, the exponent found numerically was suggested to be consistent with the $\phi^3$ theory~\cite{Liang16}. A systematic study of the $\phi^3$ theory is thus desirable.

Here, we propose an idea that the instability point is reached when the time scale of the nucleation and growth matches that of the driving arising from a temporally varying externally applied field. Integrating the theory of nucleation and growth with the $\phi^3$ RG theory of scaling for FOPTs, we are then able to construct a finite-time scaling (FTS) ansatz for the magnetization. It is found to describe remarkably well the numerical simulations of the 2D Ising model with universal instability exponents and scaling functions for two simulated temperatures below $T_c$ when allowing for a single additional universal logarithmic factor. This offers unambiguous evidence for the $\phi^3$ theory. Because the scaling form contains all the essential elements of nucleation and growth including the Boltzmann factor, which is the origin of the large discrepancy between nucleation theories and experiments, it also provides a method to rectify it.

In the following, we first review briefly the $\phi^3$ RG theory of scaling for FOPTs, the theory of FTS, and the theories of nucleation and growth to set the stage in Sec.~\ref{review}. Next, the theory of the competition between nucleation and growth and scaling with the resultant FTS ansatz is presented in Sec.~\ref{theory}. Then, after the introductions of the model, parameters and methods in Sec.~\ref{methods}, we show first in Sec.~\ref{alone} that nucleation and growth alone cannot account for the varying-field-driven transitions in the 2D Ising model below its $T_c$ and the recently proposed logarithmic time factor does not work in the model either. Then, in Sec.~\ref{results}, we verify that the scaling ansatz accounts for the numerical results well. Finally, conclusions are drawn in Sec.~\ref{summary}.

\section{\label{review}Brief review of theories}
In this section, we review briefly the $\phi^3$ RG theory of scaling for FOPTs, the theory of FTS, and the theories of nucleation and growth. One can see from the first two subsections, Secs.~\ref{phi3} and \ref{refts}, that the scaling forms with the instability exponents are identical with the corresponding ones for critical phenomena except for a finite instability point. This illustrates the unification of the two kinds of phase transitions in the present approach.

\subsection{\label{phi3}The $\phi^3$ RG theory for scaling in FOPTs}
Consider the usual $\phi^4$ model with a conventional Ginzburg-Landau functional~\cite{Wilson,Ma}
\begin{equation}
{\cal H}[\phi ] = \int {d{\rm {\bf x}}\left\{ {\frac{1}{2}\tau_4\phi ^2
+ \frac{1}{4!}g\phi ^4 + \frac{1}{2}[\nabla \phi]^2 - H\phi }
\right\}} \label{H}
\end{equation}
of a scalar-order parameter $\phi$ in the presence of an external field $H$, where $\tau_4$ is a reduced temperature and $g$ is a coupling constant that is positive for stability. It is well-known that this model possesses a critical point at $\tau_4=0$ and $H=0$ in the mean-field approximation in which $\phi$ is spatially uniform. Upon considering fluctuations, the critical point shifts to $\tau_4<0$ and $H=0$. The critical behavior of this $\phi^4$ model is found from the RG theory~\cite{Wilson,Ma} and describes the Ising universality class near its critical point at $T_c$. For a sufficiently negative $\tau_4$ or $T$ below $T_c$, there exists an FOPT at $H=0$ between two phases with opposite spatially uniform magnetization $M_{\rm eq}$. Accordingly, to study the FOPT, it is essential to shift the order parameter by $M_s$, the meaning of which will become clear shortly. In particular, let
\begin{equation}
\phi=M_s+\varphi.\label{pmvp}
\end{equation}
Then,
\begin{equation}
{\cal H} [\varphi] =
\int d{\bf r} \left[
\frac{1}{2}\tau\varphi^2 +\frac{1}{3!}gM_s\varphi^3 + \frac{1}{2}(\nabla \varphi)^2 -h\varphi\right],\label{ha}
\end{equation}
where
\begin{equation}
\tau=\tau_4+\frac{1}{2}gM_s^2, \ \ \ h= H-\tau_4M_s-\frac{1}{3!}gM_s^3, \label{tauh}
\end{equation}
and we have only kept terms containing $\varphi$ but neglected the quartic term in $\varphi$ in comparison to the cubic term in the vicinity of $M_s$. One can easily convince oneself that $\tau=0$ and $h=0$, or $M_s=\pm\sqrt{-2\tau_4/g}$ and $H_s=2M_s\tau_4/3$ from Eq.~(\ref{tauh}), is the spinodal point of the mean-field $\phi^4$ theory, the point at which the mean-field free-energy barrier for a metastable well vanishes. More importantly, it is evident by comparing Eqs.~(\ref{H}) and (\ref{ha}) that this spinodal point is the mean-field ``critical'' point of the $\phi^3$ Hamiltonian~(\ref{ha}), in exact analogy to the fact that $\tau_4=0$ and $H=0$ is the mean-field critical point of the $\phi^4$ theory~\cite{Zhongl05}. Noticing this similarity, one can thus assume that fluctuations again shift the mean-field spinodal point to some finite values, which are again denoted as $M_s$ and $H_s$ and are dubbed instability point, because this ``critical point'' is in fact an unstable point for the $\phi^3$ theory. It is then routine to perform an RG analysis to the theory, arriving at the infrared-stable fixed points and the ensuing scaling for the averaged $\phi$, or $M$, viz.,
\begin{equation}
M(H,t)=M_s+(H-H_s)^{1/\delta}f_1\left(t(H-H_s)^{\nu z/\beta\delta}\right)\label{mheq}
\end{equation}
near the instability point, where $t$ is the time, $f_1$ is a universal scaling function, and $\nu$, $\delta$, $\beta$, and $z$ are the instability exponents for the correlation length, the magnetic field, the magnetization, and the dynamics, respectively, each corresponding exactly to its critical counterpart~\cite{Zhongl05,zhong16}. Note the finite $H_s$ and $M_s$ representing the finite instability point in the scaling form different to the critical phenomena~(\ref{mheq}).

From Eq.~(\ref{mheq}), one finds an equilibration time $t_{\rm eq}\propto (H-H_s)^{-\nu z/\beta\delta}$, which diverges at the instability point. This is the origin of the scaling for the $\phi^3$ theory. For a real FOPT, however, the divergence is blocked by the neglected $\varphi^4$ term in Eq.~(\ref{ha}). Nevertheless, the $\varphi^4$ term has been shown to be irrelevant to the $\phi^3$ fixed point in the RG sense~\cite{Amit77,Elderfield,kirkham}, similar again to the critical phenomena in which all terms with orders higher than four are irrelevant~\cite{Wilson,Ma}.

The most important difference of the $\phi^3$ theory to the $\phi^4$ theory is that its nontrivial fixed points are imaginary in values and are thus usually considered to be unphysical. However, as pointed out in Sec.~\ref{intro}, the imaginariness arises from the instability of the system at just the instability points once the degrees of freedom that need finite free energy costs for nucleation are coarse-grained away. As a result, analytical continuation is necessary~\cite{Zhonge12}. This then puts the FOPT in the same universality class as the Yang-Lee edge singularity~\cite{Fisher78}, the singularity of the distribution of the Yang-Lee zeroes above $T_c$~\cite{zhong16}. This in turn enables us to improve the theoretical estimates of the static instability exponents of the $\phi^3$ theory to a three-loop order~\cite{zhong16}. In addition, these exponents agree with a functional RG calculation~\cite{li11}.

\subsection{\label{refts}Finite-time scaling}
Crucial in our analysis is the theory of FTS~\cite{Gong,Gong1,Huang,Feng}, whose essence is a constant finite time scale originating from a linear driving. This single externally imposed time scale enables us to probe effectively a process in which a system takes a long time to equilibrate, as is the present case of nucleation and growth.

To see this, let's change the field linearly with time such that $H=H_s+Rt$ with a constant rate $R$. This imposes a relationship between $H$ and $t$. Upon replacing $H$ by $R$ in Eq.~(\ref{mheq}), one finds an externally imposed time scale $t_R$
\begin{equation}
t_R=\zeta_RR^{-z/r},\label{tr}
\end{equation}
where $\zeta_R$ is a proportional coefficient independent on $R$ and
\begin{equation}
r=z+\beta\delta/\nu\label{rzbdn}
\end{equation}
is the RG eigenvalue of $R$~\cite{Zhongl05,zhong16}. By varying $R$, one then has a series of time scales at hand. When $t_{R}$ is shorter than $t_{\rm eq}$, the system falls out of equilibrium. As a consequence, the system is controlled by the driving and exhibits FTS, similar to its spatial counterpart, finite-size scaling, in which a system has a size smaller than its correlation length.

To find the FTS form, one replaces $t$ in Eq.~(\ref{mheq}) with $R$ and can then write (\ref{mheq}) in an FTS form as~\cite{Zhongl05,zhong16},
\begin{equation}
M(H,R)=M_s+R^{\beta/r\nu}f\left((H-H_s)R^{-\beta\delta/r\nu}\right),\label{ftsu}
\end{equation}
where $f$ is another universal scaling function. Equation~(\ref{ftsu}) describes the FTS regime in which $(H-H_s)R^{-\beta\delta/r\nu}\ll1$, or $t_R\ll t_{\rm eq}$ as expected. However, even when the equilibration time becomes shorter than $t_R$ and the system crosses over to the (quasi-)equilibrium scaling regime governed by Eq.~(\ref{mheq}) with $t$ replaced by $R$, Eq.~(\ref{ftsu}) can still well describe the situation~\cite{Huang,Feng}. This is because both regimes are controlled by the same fixed point.

Similarly, if we consider a finite system with a lateral length $L$, we have one more argument $L^{-1}R^{-1/r}$ in $f$ in Eq.~(\ref{ftsu}), a subleading term which constitutes a perturbation for $L\gg R^{1/r}$, viz., the driving length scale shorter than the system size, in the FTS regime.  On the other hand, if $R$ is small and/or $L$ is short such that $L\ll R^{1/r}$, the system crosses over to the finite-size scaling regime in which the corresponding term $L^rR\ll1$ becomes a perturbation~\cite{Gong,Gong1,Huang}.

\subsection{\label{nulgrow}Theories of nucleation and growth}

\subsubsection{\label{ft}FT: Field theoretically corrected theory of nucleation}
We first review briefly FT~\cite{Sides98,Sides99}. Let us start with the capillarity approximation~\cite{Gunton83} for a nucleus of an effective radius $\rho$ whose volume is assumed to be $V(\rho)=\Omega_d\rho^d$ through a shape factor $\Omega_d(T)$ in a $d$ dimensional space. The free-energy cost is then~\cite{Zia,Rikvold94}
\begin{equation}
F(\rho)=d\Omega_d^{1/d}V^{(d-1)/d}\sigma_0-V\delta\mu= d\Omega_d\rho^{d-1}\sigma_0-\Omega_d\rho^d\delta\mu,\label{fr}
\end{equation}
where $\sigma_0$ is the surface tension along a primitive lattice vector and~\cite{Harris84,Rikvold94}
\begin{equation}
\delta\mu=2M_{\rm eq}H\label{dnu}
\end{equation}
is the difference in the bulk free-energy density between the metastable and the stable phases arising from applying a positive $H$ to a state with a negative spontaneous equilibrium magnetization $-M_{\rm eq}$. Accordingly, the critical radius is
\begin{equation}
\rho_c=\frac{(d-1)\sigma_0}{\delta\mu}=\frac{(d-1)\sigma_0}{2M_{\rm eq}H},\label{rc}
\end{equation}
which maximizes $F(\rho)$, and the free-energy cost for the critical nucleus is
\begin{equation}
F_c=\Omega_d\frac{(d-1)^{d-1}\sigma_0^d}{\delta\mu^{d-1}}=\Omega_d\sigma_0^d\left(\frac{d-1}{2M_{\rm eq}}\right)^{d-1}\frac{1}{H^{d-1}}.\label{fc}
\end{equation}

For the 2D Ising model, the field-theoretically corrected nucleation rate $I(T,H)$ per unit time and volume is given by~\cite{Langer67,Gunther}
\begin{equation}
I=B(T)H^K\exp\left(-\frac{F_c}{k_{\rm B}T}\right)= B(T)H^K\exp\left(-\frac{\Xi}{H}\right)\label{ith}
\end{equation}
with $\Xi=\Omega_2\sigma_0^2/2M_{\rm eq}k_{\rm B}T$ using Eq.~(\ref{fc}), where $B(T)$ is a parameter to be adjusted to fit the numerical results, $K=3$ for the 2D kinetic Ising model~\cite{Langer67,Gunther,Gunther94,Rikvold94,Harris84}, and $k_{\rm B}$ is the Boltzmann constant.

\subsubsection{\label{bd}BD: Field-theoretically corrected Becker-D\"{o}ring theory of nucleation}
BD is based on the Becker-D\"{o}ring theory of nucleation~\cite{Becker},
\begin{equation}
I_{\rm BD}=f_c^+Z\exp\left(-\frac{F_c}{k_{\rm B}T}\right),\label{ibd}
\end{equation}
which has been found to predict the nucleation rate for the 2D Ising model very accurately without adjustable parameters if a correct free energy $F(n)$ of a droplet with $n$ spins is used~\cite{Ryu}, where the attachment rate of a molecule to the critical nucleus of $n_c$ spins is approximated as
$f_c^+\approx2\sqrt{\pi n_c}\exp(-\sigma_{\rm eff}/k_{\rm B}T)$~\cite{Ryu} and the Zeldovich factor~\cite{Zeldovich} is defined as
$Z\equiv\sqrt{\eta/2\pi k_{\rm B}T}$ with $\eta\equiv-\partial^2F(n)/\partial n^2|_{n=n_c}$ and $\sigma_{\rm eff}(T)$ denoting an effective surface tension that produces accurate values of the total interfacial free energy of a nucleus~\cite{Shneidman99}. For the 2D Ising model~\cite{Shneidman99,Zia,CNYang},
\begin{equation*}
\sigma_{\rm eff}(T)=\frac{4T}{\sqrt{\pi M_{\rm eq}}}\left\{\int_{\frac{1}{T_c}}^{\frac{1}{T}}\!\left[1-\frac{3-e^{-4x}}{\sinh(4x)}\right] \!K'[q(x)]dx\right\}^{\frac{1}{2}},
\end{equation*}
\begin{equation}
\sigma_0(T)=2+T\ln[\tanh(1/T)],\nonumber
\end{equation}
\begin{equation}
M_{\rm eq}(T)=\left[1-\sinh^{-4}(2/T)\right]^{1/8},\label{sigmaeff}
\end{equation}
where $K'$ is the elliptic integral and $q(T)=8[\cosh(4T)-1]/(\cosh(4T)+1)^2$.

In BD, the droplet free energy that can produce a correct critical nucleus is supposed to be~\cite{Ryu}
\begin{equation}
F(n)=2\sqrt{\pi n}\sigma_{\rm eff}(T)-2Hn+\tau k_{\rm B}T\ln n+c(T),\label{fbd}
\end{equation}
although the field-theoretic logarithmic correction can also be written as a preexponential factor as in Eq.~(\ref{ith}), where $\tau=5/4$ and $c(T)$ is a constant determined by matching $F(n)$ with its exact values for small $n$~\cite{Shneidman99,Ryu}.
Comparing the two forms of the surface energy in Eqs.~(\ref{fr}) and (\ref{fbd}), we find
\begin{equation}
\Omega_2(T)=\pi M_{\rm eq}\left(\frac{\sigma_{\rm eff}}{\sigma_0}\right)^2\label{ss}
\end{equation}
using $M_{\rm eq}=n/V$. As a result,
\begin{equation}
\Xi=\frac{\pi\sigma_{\rm eff}^2}{2k_{\rm B}T}.\label{xi}
\end{equation}
Equation~(\ref{fbd}) leads to both the critical nucleus
\begin{equation}
n_c=\frac{\pi\sigma_{\rm eff}^2}{16H^2}\left(1+\sqrt{1+\frac{4\tau}{\Xi}H}\right)^2,\label{nc}
\end{equation}
which matches Eq.~(\ref{rc}) for no correction $\tau=0$, confirming the good approximation, Eq.~(\ref{dnu}), used there, and, using Eq.~(\ref{ibd}), a nucleation rate in the form of Eq.~(\ref{ith}) with $K=2\tau+1/2=3$ correctly and
\begin{equation}
B(T,H)=
\frac{2^{\frac{1}{2}+3\tau}e^{\tau-(c+\sigma_{\rm eff})/k_{\rm B}T}}{\Xi^{\tau}(k_{\rm B}T)^{\frac{1}{2}+\tau}} \frac{\sqrt{y}e^{(1-y)\Xi/2H}}{\left(1+y\right)^{\frac{1}{2}+2\tau}},\label{bth}
\end{equation}
which is now $H$ dependent and we thus differentiate it from $B(T)$ by the arguments (note, however, that this differentiation is not valid for all other parameters and variables), where $y\equiv\sqrt{1+4\tau H/\Xi}$. $B(T,H)$ drops from
\begin{equation}
B(T,0)=\frac{2^{\tau}}{\Xi^{\tau}(k_{\rm B}T)^{\frac{1}{2}+\tau}}\exp\left(\tau-\frac{c+\sigma_{\rm eff}}{k_{\rm B}T}\right)
\end{equation}
to vanishing values at large $H$ monotonically.

\subsubsection{\label{avrami}Avrami's theory of growth}
\emph{In the MD regime} in which many droplets nucleate and grow~\cite{Rikvold94}, Avrami's growth law~\cite{Avrami} gives the magnetization $M$ at time $t$ as~\cite{Avrami,Sides99,Ramos}
\begin{equation}
\frac{M_{{\rm eq}H}-M(t)}{M_{{\rm eq}H}+ M_{\rm eq}} =\exp\left\{-\Omega_d\int_0^tI\left[\int_{t_n}^tv(t')dt'\right]^ddt_n\right\}\label{mta}
\end{equation}
for a system initially equilibrated in $-M_{\rm eq}$, where $M_{{\rm eq}H}(T)$ stands for the equilibrium magnetization at $H$ and $v(t)$ is the interface velocity of a growing droplet. $v\approx u H^\theta$ with $\theta=1$ and a constant proportionality $u$ in the Lifshitz-Allen-Cahn approximation~\cite{Lifshitz,Lifshitz1,Gunton83}.

For a constant external field $H$, $I$ and $v$ are constants and Eq.~(\ref{mta}) becomes
\begin{equation}
M(H,T,t)=M_{{\rm eq}H}-(M_{{\rm eq}H}+ M_{\rm eq})\exp\left[-(t/t_0)^{d+1}\right]\label{mt}
\end{equation}
in the MD regime, with a nucleation and growth time scale
\begin{equation}
t_0(H)=\zeta_0H^{-\frac{K+d}{d+1}}\exp\left\{\frac{\Xi}{(d+1)H^{d-1}}\right\},\label{t0}
\end{equation}
where $\zeta_0$ is a coefficient. Using Eq.~(\ref{ith}) and $v(t)$ from the nucleation and growth theory, we find $\zeta_0(T)=[\Omega_du^dB(T)/(d+1)]^{-1/(d+1)}$, a temperature-dependent constant. However, we will regard it as a new parameter when we consider scaling in the following. We will come back to it later on.

For a time-dependent field $H(t)=Rt$ (note that this form of field is used in the study of hysteresis only), by assuming an adiabatic approximation in which the constant field is simply replaced with its time dependent one~\cite{Sides99}, Eqs.~(\ref{ith}) and (\ref{mta}) then result in
\begin{equation}
\frac{1}{x^4}\Gamma(-4,x)- \frac{2}{x^2}\Gamma(-6,x)+ \Gamma(-8,x)=\frac{4R^3\ln2}{\Omega_2u^2B(T)\Xi^8}\label{hsx}
\end{equation}
with $x\equiv\Xi/H_c$ in two dimensions for the coercivity $H_c$ at which $M=0$, where $\Gamma$ is the incomplete gamma function and we have simply set $M_{{\rm eq}H}/(M_{{\rm eq}H}+ M_{\rm eq})=1/2$, a good approximation since $M_{{\rm eq}H}$ deviates from $M_{\rm eq}$ only slightly. In fact, we can even simply replace $M_{{\rm eq}H}$ by $M_{\rm eq}$ without appreciable difference~\cite{Zhong17}.

An identical equation with Eq.~(\ref{hsx}) has been derived for a sinusoidal driving in the low frequency approximation~\cite{Sides99} in which
\begin{equation}
R=H_0\omega=\frac{2\pi H_0}{\tau(H_0,T)R_0}\label{r0}
\end{equation}
with $\tau(H_0,T)$ being the average lifetime of the metastable state at $H_0$ and $T$~\cite{Sides99}. This is because $H=H_0\sin(\omega t)\approx H_0\omega t=Rt$ for low frequencies $\omega$. Note that $R_0$ is a scaled period~\cite{Sides99} inversely proportional to the rate $R$.
However, the sinusoidal driving has generally two controlling parameters, the field amplitude $H_0$ and the frequency $\omega$, and thus may complicate and conceal the essence of a process. In particular, at a fixed $H_0$, for $\omega\rightarrow 0$, the hysteresis loop area enclosed in a driving cycle is governed by $H_0\omega$, which is equivalent to $R$, Eq.~(\ref{r0}), and increases with $\omega$; while for $\omega\rightarrow\infty$, the area is determined by $H_0^2/\omega$ in the mean-field approximation and vanishes~\cite{Rao}. At least these two mechanisms compete and produce an area maximum at some $\omega$~\cite{Rao,Char,Sides98,Sides99}. In addition, for high $\omega$, the hysteresis loops are rounded and even drifting and thus their areas are not well defined~\cite{Sides98} because of a dynamical transition~\cite{Rao,Char,Sides99}. This subtlety does not appear in the linear driving~\cite{Zhonge}.

\emph{In the SD regime}~\cite{Rikvold94}, by neglecting the growth time for a supercritical nucleus to occupy half the system volume $L^d$ compared with the nucleation time, the probability for the system to make the transition by time $t$ is~\cite{Sides98}
\begin{equation}
P(t)=1-\exp\left[-L^d\int_0^tI(T,H)dt\right].\label{pt}
\end{equation}
Accordingly, $H_c$ is approximately determined by the time $t_c$ at which $P(t_c)=1/2$. Using again the adiabatic approximation for $I$, one obtains in this regime in two dimensions~\cite{Sides98}
\begin{equation}
\frac{1}{x^4}\Gamma(-4,x) =\frac{R\ln2}{B(T)L^2\Xi^4}.\label{hssd}
\end{equation}

An asymptotic form
\begin{equation}
H_s\sim\left[-\ln \left(\frac{\ln2}{B(T)L^2\Xi^4}R\right)\right]^{-1}\label{asymp}
\end{equation}
from Eq.~(\ref{hssd}) could be found by expanding $\Gamma(a,x)$ in large $x$~\cite{Sides98,Sides99}. This was argued to be the leading behavior for small $R$~\cite{Thomas}. However, it has been shown that such a behavior if exists could only be detected for extremely low $R$~\cite{Sides98,Sides99}, as seen by the curves marked asymptotic logarithm in Fig.~\ref{hsr}(a) below.

For BD, $H_c$ in the MD and SD regimes can be found from Eqs.~(\ref{mta}) and (\ref{pt}), respectively, though Eqs.~(\ref{hsx}) and (\ref{hssd}) are invalid.

\section{\label{theory}Theory of competition between scaling and nucleation and growth}
\subsection{Instability points}
A system lying in metastable states exhibits strong fluctuations. On the one hand, these fluctuations can lead to nucleation and growth. On the other hand, according to the $\phi^3$ theory, the fluctuations are governed by the $\phi^3$ fixed points and must show scaling and universality. Whether nucleation and growth or scaling dominates depends on their time scales. This indicates that the point at which the two time scales equal plays a pivotal role.

Our central idea is thus that scaling can be observed around the field $H_s$ that satisfies
\begin{equation}
t_0(H_s)=t_R.\label{t0tr}
\end{equation}
$H_s$ so defined divides therefore regimes in which either nucleation and growth or the scaling governed by the $\phi^3$ fixed points is dominant and hence is identified with the instability point of the theory, which was originally suggested to separate nucleation and growth from spinodal decomposition. From Eqs.~(\ref{tr}) and (\ref{t0}), Eq.~(\ref{t0tr}) becomes
\begin{equation}
H_s^{d-1}(-\ln R+\kappa \ln H_s+b)=r\Xi/[(d+1)z],\label{hs}
\end{equation}
with
\begin{equation}
\kappa\equiv r(K+d)/[z(d+1)],\qquad b\equiv r\ln(\zeta_R/\zeta_0)/z.\label{kb}
\end{equation}
In the following, we only consider FT in which $\zeta_0$ does not depend on $H$, since we will see that it is already quite good in the MD regime. How BD works in this theory will be left to future study. The corresponding $M_s$ is given by the magnetization at $H_s$ obtained from Eq.~(\ref{mt}) with $t$ replaced by $H/R$, i.e.,
\begin{equation}
M_s=M_{{\rm eq}H}-(M_{{\rm eq}H}+M_{\rm eq})e^{-H_s^{d+1}[Rt_0(H_s)]^{-(d+1)}}.\label{ms}
\end{equation}
Note that $t_R$ was obtained from $H=H_s+Rt$, or $H=H_s$ at $t=0$, while in Eq.~(\ref{mt}), $H=0$ at $t=0$. To be consistent, for $H=H_s+Rt$, the $t$ in Eq.~(\ref{mt}) must change to $t+H_s/R$, which is simply $H/R$, as we have used in Eq.~(\ref{ms}).

Several remarks are in order. First, the instability points so obtained depend on the rate $R$. This is reasonable since they rely on the probing scales as previous studies have shown~\cite{Binder78,Kawasaki,Kaski}. Only in the case in which the first two terms on the left-hand side of Eq.~(\ref{hs}) can be neglected can one arrive at a constant $H_s$. Second, as $R\rightarrow0$, one has $H_s\rightarrow0$ and $M_s\rightarrow M_{{\rm eq}H}\rightarrow M_{\rm eq}$, viz., the equilibrium transition point and magnetization, respectively, rather than the mean-field spinodal since the range of interactions is short. This is again reasonable in view of the new physical meaning of the instability point; because, at the equilibrium transition point, only nucleation and growth is possible though the transition may take a time longer than the age of the universe. Note that $M_s\rightarrow M_{\rm eq}$ instead of the initial state $-M_{\rm eq}$ as $R\rightarrow0$ since nucleation and growth have been considered. Third, if the second term on the left-hand side of Eq.~(\ref{hs}) are neglected for sufficiently low $R$, we find~\cite{Thomas} $H_s\sim (b-\ln R)^{-1/(d-1)}$, which is consistent with Eq.~(\ref{asymp}) and vanishes only for so extremely low $R$ that is not feasible numerically or experimentally~\cite{Sides99,Zhongc}. Fourth, the recently found logarithmic time factor~\cite{Pelissetto16} should be an approximated form of~(\ref{hs}), as the scaling found there is peculiar~\cite{cumulant}. So should those found numerically in Ref.~\cite{Zhongc}.

\subsection{Scaling ansatz}
Because the instability points are determined by Eqs.~(\ref{hs}) and (\ref{ms}), it is then more than natural to postulate that the scaling form~(\ref{ftsu}) changes to the ansatz
\begin{equation}
Y(H,T,R)= R^{\beta/r\nu}f\left(XR^{-\beta\delta/r\nu}(-\ln R)^{-3/2}\right),\label{ftsf}
\end{equation}
with $X(H,T,R)\equiv H^{d-1}(-\ln R+\kappa \ln H+b)-r\Xi/[(d+1)z]$ and $Y(H,T,R)\equiv M(H,T,R)-M_{{\rm eq}H}+( M_{{\rm eq}H}+M_{\rm eq})\exp\{-\zeta_0^{-(d+1)}H^{K+2d+1}R^{-(d+1)}e^{-\Xi/H^{d-1}}\}$. In~(\ref{ftsf}), we have included a logarithmic factor with an exponent $-3/2$. It may stem from either the $\phi^3$ theory in two dimensions or the neglected higher order terms in the nucleation rate~\cite{Gunther}. At present, we have no definite theory to explain it. However, we find that this single factor is sufficient for good scaling collapses for the two temperatures we simulated. It is thus universal for at least the Ising model in two dimensions.

The scaling ansatz~(\ref{ftsf}) appears complicated. However, the seemingly complicated forms of $Y$ and $X$ just reflect the competition of the nucleation and growth with the scaling. In fact,
from the scaling form~(\ref{ftsf}), at $X=0$, one recovers naturally $H_s$ that obeys Eq.~(\ref{hs}) and the magnetization satisfies
\begin{equation}
M(H_s,T,R)=M_s(T,R)+ R^{\beta/r\nu}f(0)\label{ftsm}
\end{equation}
similar to the one obtained from~(\ref{ftsu}), though $M_s$ defined in Eq.~(\ref{ms}) is rate dependent. These similarities with Eq.~(\ref{ftsu}) support our ansatz. However, owing to the competition, there is a new feature. At $Y=0$, one can only find $X|_{Y=0}=aR^{\beta\delta/r\nu}(-\ln R)^{3/2}$ for $f(a)=0$, different from the usual simple form $H|_{M=M_s}=H_s+aR^{\beta\delta/r\nu}$ obtained from Eq.~(\ref{ftsu}) at $M=M_s$.

\section{\label{methods}Model, parameters, and methods}
\subsection{model}
In order to compare with the theories, we consider the 2D Ising model with a usual Hamiltonian
\begin{equation}
{\cal H}=-J\sum_{<i,j>}s_is_j+H\sum_is_i
\end{equation}
for $L^2$ spins $s_i=\pm1$ on a square lattice. We set $J/k_{\rm B}=1$ as an energy unit. Evolution of the system involves updates of randomly selected spins from an initial state with all $s_i=-1$ in the time unit of one Monte Carlo step per spin~\cite{Binder}. For direct comparison with the results in~\cite{Sides99}, we utilize the same attempted spin flip probability $\exp(-\Delta E_i/k_{\rm B}T)/[1+\exp(-\Delta E_i/k_{\rm B}T)]$ with $\Delta E_i$ being the energy change when only the $i$th spin is flipped, work at the same $T=0.8T_c=1.815348$ and a lattice size of $L=64$ with periodic boundary conditions, and run the sinusoidal driving with the same $H_0=0.3$ and various $R_0$. However, a larger lattice size of $L=265$ is also used. For the linear driving, we started with a field to ensure equilibration sufficiently far away from the transition. This field was checked to have no effect on the results and to guarantee $M=M_{eq}$ at $H=0$ for sufficiently large $R_0$ as expected. $40$ or so periods were run for each $R_0$ in the sinusoidal driving but $25,000$ to $20,000,000$ samples were used in the linear driving for averages. For $L=256$, the largest $R_0=2000$ had the least $50,000$ samples.

\subsection{Parameters}
There are a lot of parameters in the theories. These parameters consist of two classes. One class is the parameters for the nucleation and growth. For $T=0.8T_c=1.815348$ in the 2D Ising model, all parameters except one in this class has been determined. They are $\tau(H_0=0.3,T)=74.5977$, $u=0.465(14)$, $B(T)=0.02048$, $\Xi=0.506192$, and $\Omega_2=3.15255$~\cite{Sides99,Sides98}, the last of which agrees well with $3.152543$ from Eq.~(\ref{ss}) using the exact results of $\sigma_0=0.745915$, $\sigma_{\rm eff}=0.764852$, and $M_{\rm eq}=0.954410$ from Eq.~(\ref{sigmaeff}). In addition, $K=3$ independent of the temperature as pointed out in Sec.~\ref{ft}. Besides, $c=5.31788$, which is determined by matching $F(n=1)$ to its exact values from droplet sizes of up to $4$~\cite{Ryu}. The only one that has not been determined is $M_{{\rm eq}H}$, the equilibrium magnetization at a constant external field $H$ at a fixed temperature. However, for $T<T_c$ and in an external field, the Ising system can be readily equilibrated. Therefore, $M_{{\rm eq}H}$ for a series of given $H$ at a given temperature can be found in independent simulations. Its value for any $H$ can then be obtained by interpolation.

There is another parameter in this class. This is $\zeta_0$ defined via Eq.~(\ref{t0}). Although, in the nucleation and growth theory, all parameters are known and thus $\zeta_0$ can also be estimated at $T=0.8T_c$, $B(T)$ was found by adjusting it to match the data without considering scaling~\cite{Sides99}. We therefore regard $\zeta_0$ as an adjustable parameter.

We have also studied another relatively low temperature $T=1/0.735\approx0.6T_c$ in order to verify the universality of the scaling. To this end, the parameters in the class of nucleation and growth needed are only $M_{\rm eq}$, $K$, and $\Xi$, the latter two crucially affect the nucleation and growth, as well as $M_{{\rm eq}H}$ and $\zeta_0$, as can be seen from Eqs.~(\ref{ftsf}) and (\ref{kb}). From Eq.~(\ref{sigmaeff}), the exact results are $\sigma_{\rm eff}=1.381326$ and $M_{\rm eq}=0.992879$. This gives $\Xi=2.2928252$ from Eq.~(\ref{ss}).

Another class of parameters is related to the scaling. This includes the instability exponents and $\zeta_R$. As pointed out, the $\phi^3$ theory for FOPTs falls into the same universality class as the Yang-Lee edge singularity~\cite{zhong16}. In 2D, we therefore have $\delta=-6$~\cite{Cardy85}. This gives $\nu=-5/2$ exactly~\cite{zhong16}. Also $\beta=1$~\cite{Fisher78,zhong16}. However, $z=1.753$ is only estimated to two loop orders~\cite{zhong16}. So, we may adjust it to find the best results.

Therefore, one sees that the scaling ansatz~(\ref{ftsf}) contains only two unknown parameters, $\zeta_0$ and $\zeta_R$ or $b$ because of Eq.~(\ref{kb}), to be determined for the 2D Ising model. All the other parameters are either known exactly or can be determined from independent sources. These few degrees of freedom of the model offer therefore an excellent arena for testing the theory.

There is a logarithmic correction term with the exponent $-3/2$. We will see that without this term, the curves of different rates cross one point correctly. With this single correction, they overlap over a large range after rescaled. In practice, we try only simple numbers for the exponent and determine it for one temperature. Then we find that the same factor also works well for the other temperature. This demonstrates its universality in the 2D model. In critical phenomena, one also encounters logarithmic corrections. However, what is the exact origin of the term has yet to be identified.

\subsection{\label{elim}The method of varied-range fitting}
In order to show whether an FTS form can really explain numerical results or not, one method is to collect data from a series of rate $R$ and then fit the data according to the scaling form. To this end, it is essential to check that the results such as exponents of the fit are independent on $R$ at least for some of its ranges. Otherwise, the exponents obtained are only apparent. To this end, we employ a method of sequentially varied range fitting.

The method works as follows. Let there be $N$ data measured at $R_1$, $R_2$, $\cdots$, $R_N$, arranged in ascending order. We first fit the six data from $R_N$ to $R_{N-5}$ and obtain an apparent exponent $A$. It can be plotted as a point at $R_{N-5}$ in the plane of $A$ versus $R$. Then, we include one more datum at $R_{N-4}$ and fit the seven data from $R_N$ to $R_{N-4}$ again and obtain another $A$. This adds another point in the plane at $R_{N-4}$. The procedure is repeated until all $R$ are included in the fit, leading to a curve in the plane. One can of course plot the points on the large $R$ side at $R_N$, $R_{N-1}$, $\cdots$, $R_5$, because the curve only demonstrates the variation of $A$ with $R$. Next, we drop the datum at the largest $R_N$ and start the series of fits from $R_{N-1}$. This results in another curve in the $A$ versus $R$ plane. Repeating the procedure until the last point for $R_1$ to $R_5$ is found. One can of course start the fits from $R_1$ to $R_5$ reversely. Anyway, the results show all possible sequential fitting to the data with at least six rates and display a series of curves with systematically different lengths. As a consequence, one can readily identify whether there exists a plateau for a rate-independent exponent.

\subsection{\label{proc}Procedure}
The procedure to verify the scaling ansatz~(\ref{ftsf}) is as follows. Given a $z$, we guess a value of $b$ and solve out $H_s$ from Eq.~(\ref{hs}), find the corresponding $M(H_s)$ for a series of $R$ from the simulated magnetization curves, and then fit them according to Eqs.~(\ref{ftsm}) and (\ref{ms}). Note that Eq.~(\ref{ftsm}) produces one more parameter, $f(0)$. Yet, because $b$ is given, we are left only with two parameters, $f(0)$ and $\zeta_0$, to be determined from the fit. However, we have also to estimate the range of $R$ that fits into the MD regime in which $t_0$ is determined. As three parameters can usually be accurately found from a nonlinear fit, we thus regard one more parameter, either $\beta/r\nu$ or $d + 1$ in Eqs.~(\ref{ftsm}) and (\ref{ms}) as unknown. In practice, we run separate fits to determine both $\beta/r\nu$ and $d + 1$ for two different temperatures in order to show that the outcomes of the fits are not just accidental and the function indeed describes the results well. Employing then the method of varied-range fitting, we find a series of apparent results, including the apparent exponent $\beta/r\nu$ or $d + 1$. The correct $b$ must lead to the right $\beta/r\nu$ or $d + 1$ for the given $z$, if the theory does describe the numerics. Moreover, those rates that give rise to the correct exponent must fall in the MD regime consistently. All the results can finally be plugged in Eq.~(\ref{ftsf}) for a corroboration. Remarkably, the two time scale coefficients can be found accordingly.

\section{\label{alone}Lack of simple scaling in nucleation and growth}
In this section, we first show numerically that classical nucleation and growth theories alone cannot explain the hysteresis of the FOPT of the paradigmatic 2D Ising model driven by linearly varying an externally applied field.
Although both FT and BD agree quite well generally with the simulation results, the slight but systematic deviations for different sweep rates $R$ of the driving indicate that the theories miss something for such a driven transition. We then show that the recently found scaling with a logarithmic time factor of $\ln^2t$ in the Potts model~\cite{Pelissetto16} does not work in the present case.

\begin{figure}[t]
\centerline{\includegraphics[width=1\linewidth,height=11.2cm]{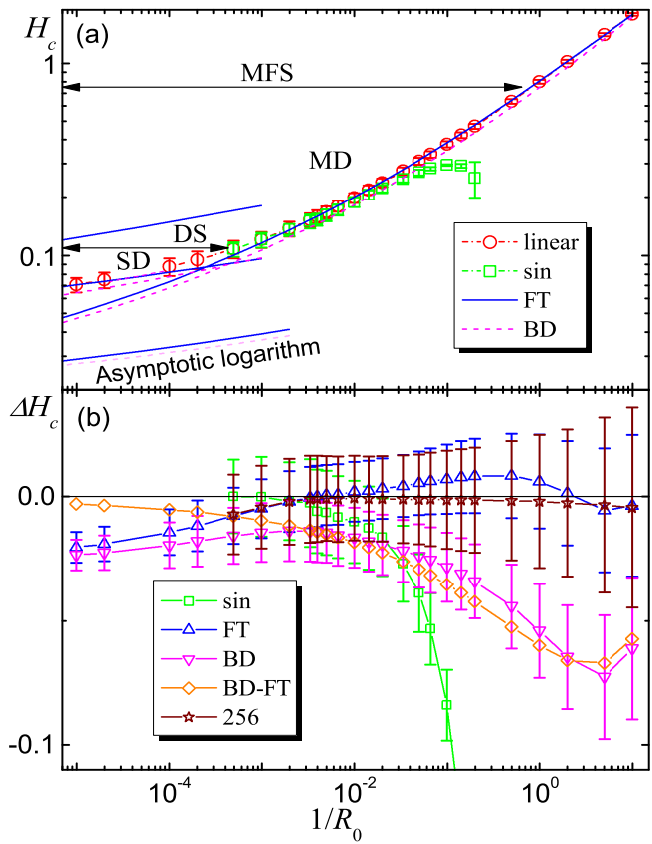}}
\caption{\label{hsr}(Color online) (a) $H_c$ versus the reciprocal of the scaled period $R_0$ for $L=64$ at $T=0.8T_c$. Linear and sin denote the data obtained from Monte Carlo simulations of the 2D kinetic Ising model using $H=Rt$ and $H=H_0\sin\{2\pi t/[\tau(H_0, T)R_0]\}$, respectively. The three curves around SD are theoretical results for the SD regime [one BD and two FT curves with $B(T)=0.02515$ for the upper and $B(T)=69.73$ for the lower] and the two lower curves are results of the asymptotic logarithmic approximation, Eq.~(\ref{asymp}) [the curve of the larger $B(T)$ is far smaller and absent]. The horizontal lines with arrows indicate the dynamic spinodal (DS, which separates regimes of MD and SD) and the mean-field spinodal (MFS, above which spinodal decomposition occurs)~\cite{Tomita,Rikvold94}. Note that the ``error bars" give the standard deviations of the distributions of the transition involved~\cite{Sides99}. (b) Differences in $H_c$. BD-FT stands for the differences of the two theories, while the others are the differences to the linear driving. $256$ symbols the differences between results on $L=64$ and $L=256$ lattices.}
\end{figure}
Figure~\ref{hsr} displays the simulation results along with theoretical ones from solving numerically Eqs.~(\ref{hsx}), (\ref{hssd}) and those for BD. One sees that the results of linear and sinusoidal drivings are identical for large $R_0$. Using the values of $H_c$ at $R_0=200$ in the linear driving, we find from Eq.~(\ref{hsx}) $B(T)=0.02515$, which is close to $0.02048$ found from the same $R_0$ for the sinusoidal driving in Ref.~\cite{Sides99} but produces slightly better results. As seen in Fig.~\ref{hsr}(a), the predictions of FT are excellent in the MD regime and even beyond, while in the SD regime, they are poor. To match the lowest rate, we find $B(T)=69.73$, larger by more than two thousand times. In contrast, BD yields good results even in the SD regime without any adjustable parameters, though they are slightly smaller as seen in Fig.~\ref{hsr}(b) and the $H$ range is far larger than $0.01$ to $0.13$ studied in Refs.~\cite{Ryu,Ryu1} for a constant $H$.

\begin{figure}
\centerline{\includegraphics[width=1\columnwidth,height=11.2cm]{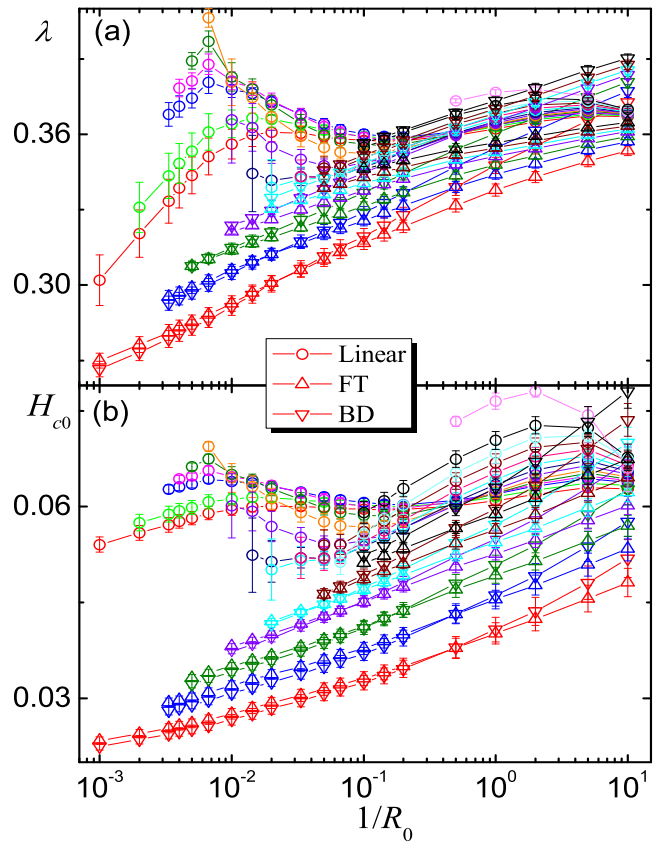}}
\caption{\label{lamhc}(Color online) Varied-range fitting results of (a) $\lambda$ and (b) $H_{c0}$ for $L=64$ at $T=0.8T_c$. The fits start from the small $R$ or large $R_0$ side and the results are plotted at the largest $R$ or smallest $R_0$ (see Sec.~\ref{elim} for details). For clarity, we plot only every other curve for the theories. Lines connecting the symbols are only a guide to the eye. The curves of identical colors have identical ranges of $R_0$. Note the trends and the large discrepancy between the theories and simulations in (b) irrespective of the dense curves which may not be easy to follow.}
\end{figure}
\begin{figure}
\centerline{\epsfig{file= 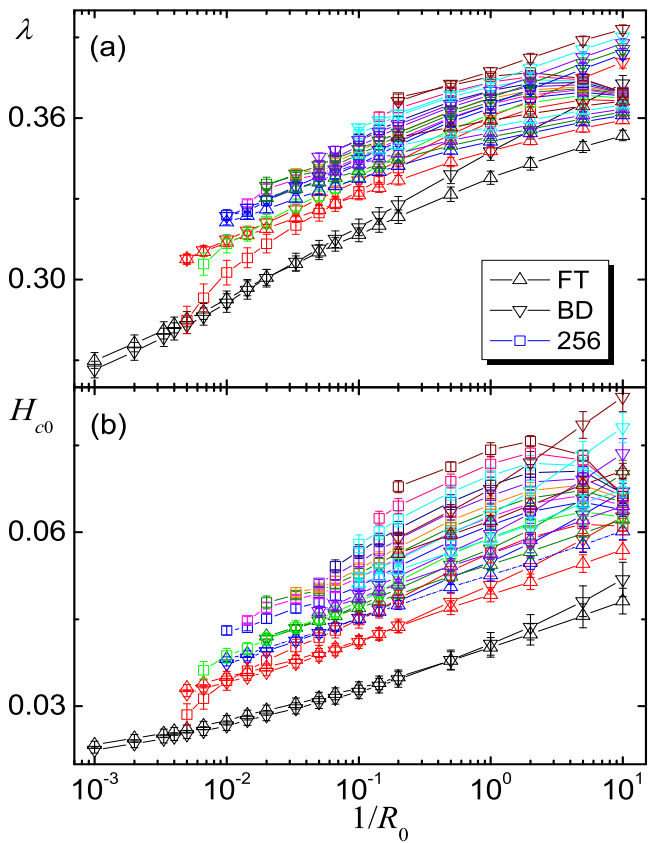,width=1\columnwidth,height=11.2cm}}
\caption{\label{rr256tw}(Color online) Varied-range fitting results of (a) $\lambda$ and (b) $H_{c0}$ for $L=256$ at $T=0.8T_c$. No apparent concave around $R_0=10$ appears. Eleven successive curves for the linear driving are shown, while every other curve for the theories are shown except for the lowest curve. Lines connecting the symbols are only a guide to the eye. The curves of identical colors have identical ranges of $R_0$.}
\end{figure}
However, from the differences shown in Fig.~\ref{hsr}(b), one sees that both theories exhibit systematic deviations from the numerical results. This can be clearly seen from Fig.~\ref{lamhc}, where we show the results of the varied-range fitting to~\cite{Zhonge,Zhonge1}
\begin{equation}
H_c=H_{c0}+cR_0^{-\lambda}\label{hcr0}
\end{equation}
with constants $H_{c0}$, $c$, and $\lambda$. For the theories, both $\lambda$ and $H_{c0}$ change monotonically with the range of $R_0$ that is used to fit them out, conforming to the expectation that the results described by such theories exhibit no scaling~\cite{Sides98,Sides99}. However, the simulation results are qualitatively distinct. This is also true for $L=256$ for which the features at large $R_0$ in Fig.~\ref{lamhc} disappear owing to the suppression of the SD regime in the range studied as seen in Fig.~\ref{rr256tw}. In Fig.~\ref{lamhc}, although $\lambda$ from the simulations appears close to that from BD in a small range (the cyan curve), it is clear that there exists a substantial discrepancy in $H_{c0}$ between the theories and the numerical results both for $L=64$ and $L=256$. Note that this large gap cannot be removed by adjusting parameters like $B(T)$, because bigger $H_{c0}$ leads to bigger $\lambda$ and thus the gap transfers to $\lambda$. Moreover, such possible adjustments have only a negligible effect because the differences in $H_c$ between the theories and the numerical results are small. Furthermore, larger lattice sizes and even the thermodynamic limit cannot remove the gap. Comparing Figs.~\ref{lamhc} and~\ref{rr256tw}, one sees that the results on the right hand side for large $R$ (small $R_0$) are similar. In fact, when $L$ is larger than the driving length scale in the FTS regime, it plays only a negligible role~\cite{Gong1}, as pointed out in Sec.~\ref{refts}. Accordingly, the results on different lattice sizes must differ negligibly in principle. Therefore, there must be something missing in the nucleation theories.

\begin{figure}
\centerline{\includegraphics[width=\linewidth]{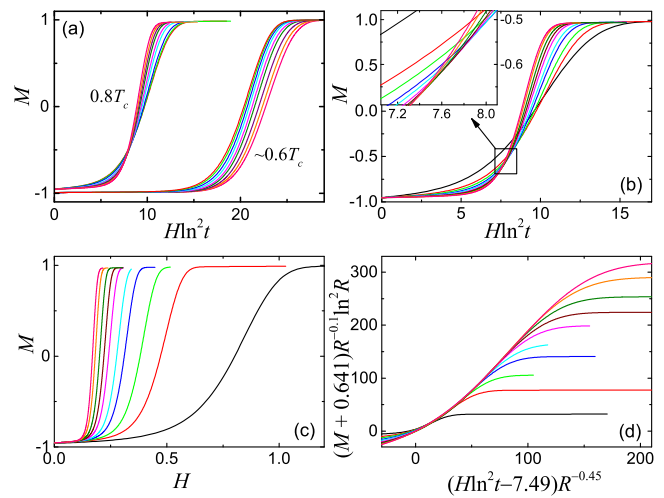}}
\caption{\label{hln2t}(Color online) (a) $M$ versus $H\ln^2\!t$ for nine $R$ from $R=0.00421$ to $0.000168$ (from right to left above the crossing) at $T=0.8T_c$ and from $R=0.00145$ to $0.000116$ (from left to right) at $T\approx0.6T_c$. (b) The crossing point at about $H\ln^2\!t=H_s=7.49$ and $M_s=-0.641$ and its enlargement at $T=0.8T_c$. (c) Original $M$ versus $H$ curves at $T=0.8T_c$. (d) Collapses of all curves in (c) including even the largest rate $R=0.00421$ for the chosen exponents.}
\end{figure}
Recently, scaling was found to emerge if a $\ln^2\!t$ term was utilized in the FOPTs of the Potts model, in which the field is served by $T-T_0$ with $T_0$ the equilibrium transition temperature~\cite{Pelissetto16}. This factor was argued to arise from the interplay between the exponential time in tunneling between the two phases and the droplet formations in the low-temperature phase. The curves of energy differences---normalized by their fixed value at $T_0$---versus $(T-T_0)\ln^2\!t$ for various cooling rates cross at a finite value. This was suggested to show a dynamic transition with spinodal-like singularity~\cite{Pelissetto16}. Figure~\ref{hln2t}(a) shows that this crossing appears to happen for the Ising model studied here at $T=0.8T_c$. However, it is absent at $T\approx0.6T_c$. This indicates that the mechanism can not be dominated generally, as varying $T$ or $H$ cannot change the mechanism and other corrections should not be so large as to substantially change the leading behavior.

\begin{figure}
\centerline{\epsfig{file=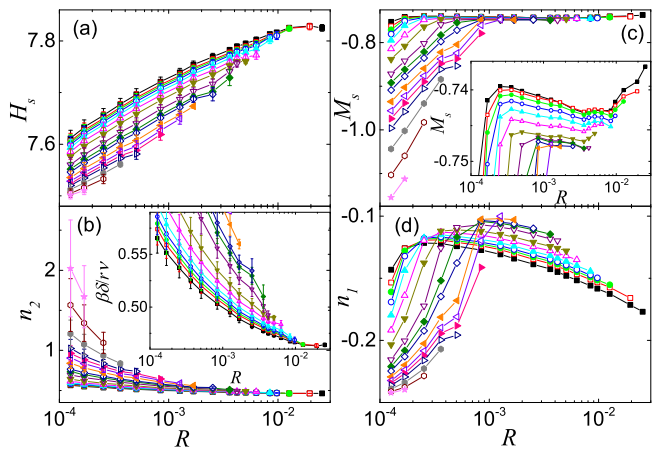,width=1\columnwidth}}
\caption{\label{y0an}(Color online) Varied-range fitting results of $n_2$ and $H_s$ and $n_1$ and $M_s$ at $T=0.8T_c$. (a) and (b) are results of the fits of the $H$ at $M_s=-0.641$ for various $R$ and (c) and (d) are those of the fits of the $M$ at $H_s=7.49$. The insets are the enlarged views of the parts that appear to be constant. Lines connecting symbols are only a guide to the eye.}
\end{figure}
Moreover, Fig.~\ref{hln2t}(b) shows that the crossing point is in fact not a single point. Yet, as illustrated in Fig.~\ref{hln2t}(d), we are able to find exponents and a logarithmic correction to collapse all the curves displayed in Fig.~\ref{hln2t}(c), including even the largest rate plotted that does not cross the point at all, as seen in Fig.~\ref{hln2t}(b). Figure~\ref{hln2t}(d) indicates that
\begin{equation}\label{mhscalingwr}
M=M_s+R^{n_1}(-\ln\! R)^{-2}f_2[(H\ln^2\!t-H_{s})R^{-n_2}],
\end{equation}
where $f_2$ is a universal scaling function. Therefore, at $M=M_s$, one must have $H\ln^2\!t=H_s+a_1R^{n_2}$, while at $H\ln^2\!t=H_s$, $M=M_s+f_2(0)R^{n_1}(-\ln\!R)^{-2}$ with $f_2(a_1)=0$. We can then perform varied-range fittings both at $M=M_s$ and at $H\ln^2\!t=H_s$ to investigate whether the corresponding exponents $n_2$ and $n_1$, respectively, are indeed constant or not. The results are shown in Fig.~\ref{y0an}. One sees from Fig.~\ref{y0an} that all results depend strongly on $R$. Moreover, there exists no consistency in $H_s$ and $M_s$, viz., the fit at $M_s$ cannot reach the $H_s$ at which the fit was performed and vice versus. Further, the exponent $n_1$ has even a wrong sign opposite to that used for the collapses in Fig.~\ref{hln2t}(d)! Note that the collapse in Fig.~\ref{hln2t}(d) becomes poor when $n_2$ varies about $\pm0.1$ or $n_1$ changes to $-0.1$. In fact, according to Eq.~(\ref{mhscalingwr}), one should have found the crossing point using $(M-M_s)R^{-n_1}(-\ln\! R)^{2}$ versus $H\ln^2\!t$ rather than $M$ versus $H\ln^2\!t$, because all curves of different $R$ could then really cross at a single point for the former. This is the usual method for finding the critical point in finite-size scaling~\cite{cumulant}. Therefore, the logarithmic time factor does not apply to the Ising model even for the case of $T=0.8T_c$.

In fact, a further logarithmic correction can lead to plateaus in the fitted results and can even change $n_1$ to the value of $\beta/r\nu$ and $n_2$ to that of $\beta\delta/r\nu$ and consistency in $H_s$ and $M_s$~\cite{Zhongc}. However, we believe this together with the $\ln^2\!t$ factor is only the approximated forms of the present theory, since a lot of parameters are needed.

\section{\label{results}Verification of the theory}
We now follow the procedure in Sec.~\ref{proc} to verify the theory of scaling near the proposed instability point by showing that it accounts for the numerical results well.

\begin{figure*}
\centerline{\includegraphics[width=0.95\textwidth]{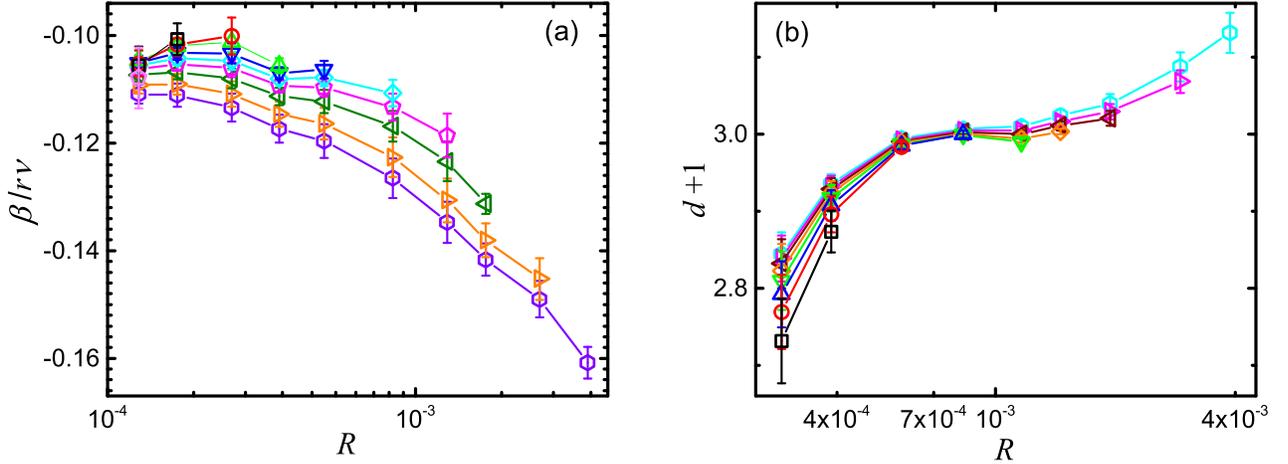}}
\caption{\label{hmexp}(Color online) Apparent exponents (a) $\beta/r\nu$ and (b) $d +1$ for $T=0.8T_c$ and $T\approx0.6T_c$, respectively, obtained from the varied-range fitting of Eqs.~(\ref{ftsm}) and (\ref{ms}). Here we start the fit from the large $R$ side and plot the results at the largest $R$, see Sec.~\ref{proc} for details. $b=2.82$ and $0.77$ for $T = 0.8T_c$ and $T \approx 0.6T_c$, respectively. Lines connecting symbols are only a guide to the eye.}
\end{figure*}
Figure~\ref{hmexp} shows the results of the varied-range fitting for $z = 1.5$ from Eqs.~(\ref{ftsm}) and (\ref{ms}). One sees that, as the data of large $R$ are omitted in the fits, the apparent exponents approach respectively $-0.103$ and $3$ correctly. In Fig.~\ref{hmexp}(b) for $T\approx0.6T_c$, including of smaller rates again drives the exponent away from $3$, a feature which we will come back later on. The rates that give rise to the correct exponents $\beta/r\nu=-0.103$ and $d +1=3$ within the error-bars are thus chosen to be the five data of the sixth curve (inverse triangles, counting from the rightmost end point) plus the five larger rates in the varied-range fitting, adding up to ten rates for $T = 0.8T_c$, and the three data of the fourth curve (diamonds) plus the five larger rates, totally eight rates for $T \approx 0.6T_c$. We have displayed different exponents for different temperatures in Fig.~\ref{hmexp} in order to show that both exponents indeed reach their correct values in contrast with Figs.~\ref{hln2t} and \ref{y0an}. The other exponent that is not displayed exhibits similar behavior and shows consistently that Eq.~(\ref{ftsm}), together with its universal instability exponents and Eq.~(\ref{ms}), can indeed account for the simulation results.

\begin{figure*}
\centerline{\includegraphics[width=\linewidth]{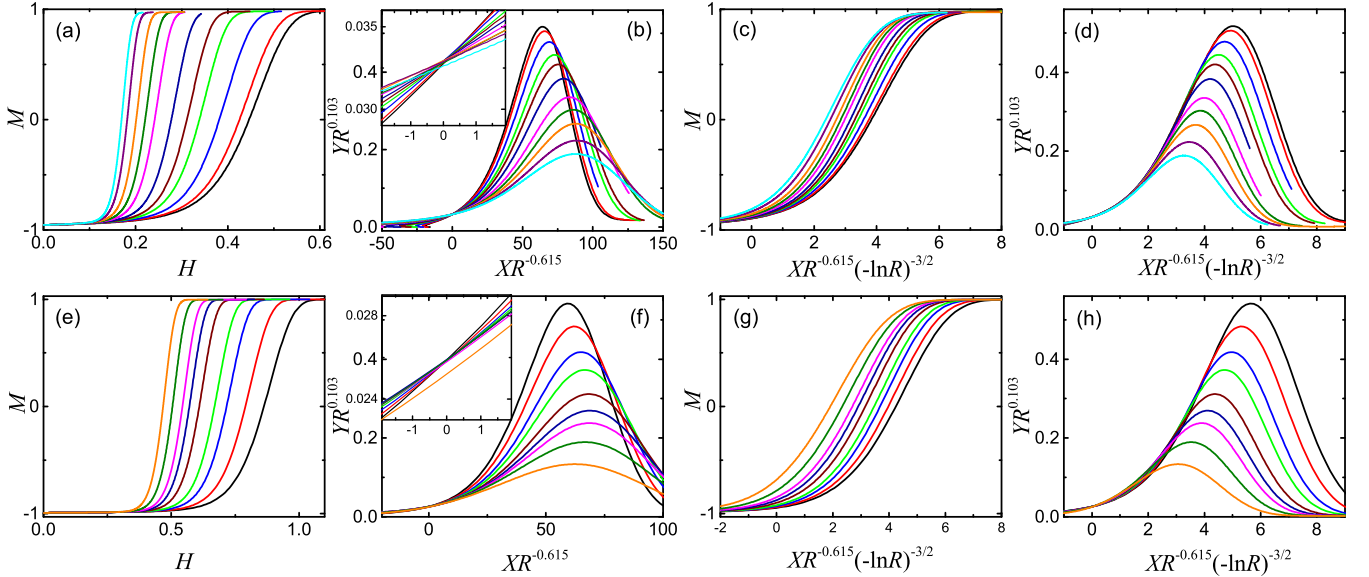}}
\caption{\label{rgt}(Color online) Magnetization curves and their various rescalings for (a) to (d) $T=0.8T_c$ and (e) to (h) $T\approx0.6T_c$. (a) and (e) are the magnetization curves for ten $R$ from about $0.00421$ to $0.000 168$ (or $R_0$ from $6$ to $150$) and eight $R$ from about $0.008 30$ to $0.000 830$, respectively. We have also shown one more small rate at $0.000126$ (or $R_0=200$) and $0.000 581$ for the two temperatures, respectively. As $R$ decreases, the curves shift to the left. (d) and (h) display the rescaled curves of all curves in (a) and (e), respectively. (c) and (g) depict the curves with only $H$ being rescaled, while (b) and (f) show the rescaled curves in the absence of the logarithmic correction. The insets enlarge the parts near the crossing points. $\zeta_0$ is $36.3$ and $59.9$ for $T = 0.8T_c$ and $T \approx 0.6T_c$, respectively. Identical colors represent curves of identical rates for the same temperature only.}
\end{figure*}

Using the fitted results of $\zeta_0$ and the ranges of rates that produce the correct exponents, we rescale the magnetization curves shown in Fig.~\ref{rgt}(a) and \ref{rgt}(e) according to the scaling form~(\ref{ftsf}) in the absence of the logarithmic correction and plot the results in Fig.~\ref{rgt}(b) and \ref{rgt}(f). One sees that at the crossing point $X=0$ at which the fits were performed, all curves cross perfectly except the two extra rates included. This is in sharp contrast with Fig.~\ref{hln2t}(b). We emphasize that these crossings are obtained by the two parameters $\zeta_0$ and $\zeta_R$ or $b$ only. Other parameters, including the two universal instability exponent composites and those for nucleation and growth, are all predetermined. Note that we have also given a fixed $z$. However, it was chosen to be around the theoretical one-loop value. Moreover, other $z$ values yield similar results. These therefore show that the results do fit the simulations well.

To collapse the other portions of the curves, we find that a single logarithmic correction with a simple exponent $-3/2$ works well for both temperatures studied. The results are depicted in Fig.~\ref{rgt}(d) and \ref{rgt}(h). The peaks of the rescaled curves stem from the competition between $M$ and the part of nucleation and growth in $Y$ and lie in the late stages of the transition as seen in Fig.~\ref{rgt}(c) and \ref{rgt}(g). One sees that the rescaled curves collapse onto each other almost perfectly even relatively far away from the instability point at $X = 0$ and even for rates beyond. Together with the exponent plateaus shown in Fig.~\ref{hmexp}, this strongly validates the scaling form. We note, however, that, in the absence the universal exponents and nucleation and growth, the simple logarithmic term alone certainly cannot collapse the curves. For example, consistent collapses in contrast with Fig.~\ref{hln2t}(d) can indeed be found numerically as mentioned in the last section, though several other logarithmic terms besides again the $\phi^3$ theory must be taken into account~\cite{Zhongc}.

Similar scaling collapses appear for $z$ bigger than $1.5$ and even up to $2.5$ plus. We choose $1.5$ because the scaling functions for the two temperatures appear nearly parallel. This can be seen in Figs.~\ref{rgt}(d) and \ref{rgt}(h), where identical scales are employed. The two rescaled curves only displace with each other by less than $0.01$ in $f(0)$.  This slight difference may result from the neglected higher order terms in the nucleation rate~\cite{Gunther}, which may also be a source of the extra logarithmic factor. As pointed out above, the rescaled curves already cross at $X = 0$ and $Y(H_s,T,R)R^{-\beta/r\nu}=f(0)$ even without the logarithmic factor, as illustrated in Figs.~\ref{rgt}(b) and \ref{rgt}(f). This indicates that at least the predominant contribution of the nucleation and growth has been taken into account in the present theory.

\begin{figure}
\centerline{\includegraphics[width=\linewidth]{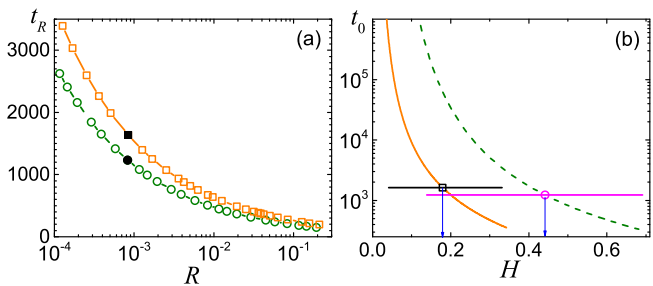}}
\caption{\label{trt0}(Color online) (a) The driving time scale $t_R$ versus $R$ and (b) the nucleation and growth time scale $t_0$ versus the field $H$ at $T = 0.8T_c$ (squares/solid line) and $T \approx 0.6T_c$ (circles/dashed line). The two horizontal lines are $t_0(H_s) = t_R$, corresponding to the two filled black symbols in (a). The two vertical arrows separate the scaling regime on the left from the nucleation and growth regime on the right. Lines connecting symbols are only a guide to the eye.}
\end{figure}
Figure~\ref{trt0} displays the time scales. From Eqs.~(\ref{tr}) and (\ref{t0}), $t_R$ (or $t_0$) increases as $R$ (or $H$) decreases and diverges as $R$ (or $H$) vanishes. However, $t_0$ rises exponentially fast than $t_R$ as seen in the figure. This implies that the two curves always intersect at a finite $H_s$ practically~\cite{Zhonge,Zhonge1} and hence there exists always an FTS regime in which the driving dominates the dynamics no matter how small $R$ is. From the dependences of $t_R$ and $t_0$ on $R$ and $H$, respectively, it is evident that $H_s$ increases with $R$. As a consequence, large rates drive the transition to take place at large fields as Figs.~\ref{rgt}(a) and (e) have demonstrated. In addition, because the free-energy cost for nucleation increases significantly as $T$ is lowered, $t_0$ increases rapidly as $T$ decreases,  though $\zeta_0$ and $\zeta_R$ only change moderately, from respectively $36.3$ and $107.4$ at $T = 0.8T_c$ to $59.9$ and $80.5$ at $T \approx 0.6T_c$, with reverse temperature dependences as Fig.~\ref{trt0} displays. Therefore, the transition occurs at a large field and the hysteresis increases as $T$ is lowered, as can also be seen in Figs.~\ref{rgt}(a) and (e).

\begin{figure}[t]
\centerline{\includegraphics[width=\linewidth]{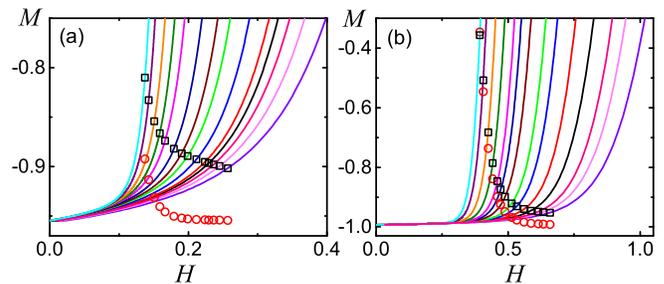}}
\caption{\label{hsms}(Color online) The instability points and the magnetization at $H_s$ for (a) $T = 0.8T_c$ and (b) for $T \approx 0.6T_c$. $M(H_s)$ (squares) lies on the magnetization curves (lines) whereas $M_s$ (circles) does not. In (a) and (b), the field rates $R$ increase from left to right. Three more largest rates are drawn for both temperatures as compared to Figs.~\ref{rgt}(a) and (e). Two more smallest rates are also plotted in (b). The color codes are identical with Fig.~\ref{rgt}.}
\end{figure}

Figure~\ref{hsms} illustrates the magnetization at $H_s$, $M(H_s)$, and the instability points at $H_s$ and $M_s$. $M(H_s)-M_s$ is just $f(0)R^{\beta/r\nu}$ from Eq.~(\ref{ftsm}). One sees that scaling and universality persist though the instability points appear somehow far away, possibly because the time scale $t_0$ is determined only by $H_s$ but not by $M_s$. It is clear that $H_s$ decreases while $M_s$ increases as $R$ is reduced as expected. A unique feature is that, for $T \approx 0.6T_c$, $M(H_s)$ and $M_s$ rise sharply for low $R$ and hence low $H_s$ and cross each other. This reveals the reasons why the small rates deviate from scaling at the temperature shown in Fig.~\ref{hmexp}(b). On the one hand, for small rates, the system has possibly already left the MD regime to which the present $t_0$ is applicable. Indeed, at low temperatures the critical nucleus is large and the system can readily enter the SD regime for small lattice sizes~\cite{Rikvold94,Tomita}. One must accordingly employ the time scale pertinent to the SD regime instead of the MD one in the theory. On the other hand, the characteristic of the SD regime is large fluctuations in the coercivity $H_c$~\cite{Tomita}. So, even in the crossover between the two regimes, the large fluctuations may bring a small shift to the field of the magnetization curve. This shift can thus give rise to a large deviation in $M(H_s)$ in comparison with the fixed theoretical $H_s$ for a given $R$.
For the large rates, their magnetizations may be too nonequilibrium to show scaling. Indeed, they are found to depend slightly on the initial state. Possibly, yet another time scale would have to be considered for these rates. All these could therefore result in the limited range of validity of the present theory developed in the MD regime.

Owing to the difference between $M(H_s)$ and $M_s$, the time scale coefficient $\zeta_0$ we have found is different from the value $5.59$ obtained from its definition at $T = 0.8T_c$. Conversely, the present value of $36.3$ leads back to $B = 0.000 092 0$, about $270$ times smaller as compared to $0.025 15$. This appears not so absurd as deviations of orders of magnitude are common in the field. For example, to fit the results at the same temperature in the SD regime, $B$ must be more than $2 000$ times bigger as mentioned in Sec.~\ref{alone}. In addition, the difference $M(H_s)-M_s$, or $f(0)R^{\beta/r\nu}$, the scaling term, should also be responsible for the gap in $H_{c0}$ between the nucleation and growth theories and the simulation results seen in Figs.~\ref{lamhc} and \ref{rr256tw} and thus should be the reason why the nucleation and growth theories alone cannot account for scaling.

\section{\label{summary}Conclusion}
We have constructed and verified a theory for the dynamics of first-order phase transitions by integrating the theory of nucleation and growth with the $\phi^3$ RG theory for dynamic scaling and universality in those transitions and the theory of finite-time scaling. The theory relies on the time scale of nucleation and growth and the time scale of driving and offers a new physical interpretation of both the instability points and the division of the scaling and the nucleation and growth regimes in the dynamics. It also indicates that the instability point against which the $\phi^3$ theory is expanded may be spatially nonuniform. On the one hand, despite being interwoven with nonuniversal nucleation and growth, scaling and universality have been unambiguously verified in the 2D Ising model below its critical temperature. As a consequence, first-order phase transitions can be studied similar to their continuous counterpart using scaling and universality and hence the theories for both kinds of transitions can be unified within the framework of the renormalization-group theory. On the other hand, the intimate relationship with scaling and universality provides a new way to accurately determine the theory of nucleation and growth.

\section*{acknowledgments}
I thanked Shuangli Fan, Xuanmin Cao, and Weilun Yuan for their useful discussions. This work was supported by National Natural Science Foundation of China (Grant No. 11575297).

\end{document}